\documentclass[journal]{IEEEtran}
\usepackage{siunitx}
\usepackage{amsfonts,amsmath,amsthm,mathtools}
\usepackage{graphicx,xspace,epsfig,syntonly,dsfont}
\usepackage{url,cite,bm,bbm}
\usepackage{algorithmicx}
\usepackage{arydshln}
\usepackage{booktabs}
\usepackage{balance}
\usepackage{stfloats}
\usepackage[caption=false]{subfig}
\usepackage{diagbox}
\usepackage{multirow}
\usepackage{enumitem}
\usepackage{subfig}

\usepackage{pifont}

\usepackage{algorithm}
\usepackage{algpseudocode}
\usepackage{graphicx}

\long\def\symbolfootnote[#1]#2{\begingroup
\def\thefootnote{\fnsymbol{footnote}}
\footnote[#1]{#2}\endgroup}
\usepackage{xcolor}
\DeclareSIUnit \voltampere {MVA} 

\IEEEoverridecommandlockouts

\allowdisplaybreaks[4]

\title{Adversarial Multi-Agent Reinforcement Learning for Proactive False Data Injection Detection}

\author{
    \IEEEauthorblockN{
        Kejun Chen\IEEEauthorrefmark{1},
        Truc Nguyen\IEEEauthorrefmark{1},
        Abhijeet Sahu \IEEEauthorrefmark{2},
        Malik Hassanaly\IEEEauthorrefmark{1}
    }
    
    \thanks{\IEEEauthorblockA{\IEEEauthorrefmark{1}Computational Science Center, National Laboratory of the Rockies (NLR), Golden, CO, USA.}}

    \thanks{\IEEEauthorblockA{\IEEEauthorrefmark{2}Cyber Security Center, NLR, Golden, CO, USA.} Emails: \texttt\{kejun.chen, truc.nguyen, bhijeet.Sahu, malik.hassanaly\}@nrel.gov. The corresponding author is Kejun Chen.}
    
    \thanks{This work was authored by the National Laboratory of the Rockies for the U.S. Department of Energy (DOE), operated under Contract No. DE-AC36-08GO28308. This work was supported by the Laboratory Directed Research and Development (LDRD) Program at the National Laboratory of the Rockies. The views expressed in the article do not necessarily represent the views of the DOE or the U.S. Government. The U.S. Government retains and the publisher, by accepting the article for publication, acknowledges that the U.S. Government retains a nonexclusive, paid-up, irrevocable, worldwide license to publish or reproduce the published form of this work, or allow others to do so, for U.S. Government purposes.
    This research was performed using computational resources sponsored by the U.S. Department of Energy's Office of Critical Minerals and Energy Innovation and located at the National Laboratory of the Rockies.}
}

\begin{document}
\maketitle

\begin{abstract}
Smart inverters are instrumental in the integration of distributed energy resources into the electric grid. Such inverters rely on communication layers for continuous control and monitoring, potentially exposing them to cyber-physical attacks such as false data injection attacks (FDIAs). We propose to construct a defense strategy against a priori unknown FDIAs with a multi-agent reinforcement learning (MARL) framework. The first agent is an adversary that simulates and discovers various FDIA strategies, while the second agent is a defender in charge of detecting and locating FDIAs. This approach enables the defender to be trained against new FDIAs continuously generated by the adversary. In addition, we show that the detection skills of an MARL defender can be combined with those of a supervised offline defender through a transfer learning approach. Numerical experiments conducted on a distribution and transmission system demonstrate that: a) the proposed MARL defender outperforms the offline defender against adversarial attacks; b) the transfer learning approach makes the MARL defender capable against both synthetic and unseen FDIAs.

\end{abstract}

\begin{IEEEkeywords}
Multi-agent reinforcement learning, transfer learning, false data injection attack detection
\end{IEEEkeywords} 

\section{Introduction}
Integration of intermittent and inverter-based energy sources makes the electric grid less resilient to frequency and voltage instabilities~\cite{denholm2020inertia}. Smart inverters are a promising solution that can help regulate voltage and frequency, thanks to actuation strategies that depend on the state of the system~\cite{tamrakar2017vsm_review}. Smart inverters rely on information exchange, either through observations of the grid's state or through actuation decisions sent via communication networks~\cite{Yuanliang}. This makes them a target for cyber-physical attacks, and in particular false data injection attacks (FDIAs) that can refer to denial-of-service attacks \cite{liang20162015}, tampering of sensor measurements~\cite{liu2011false} or even modifying the control logic of the inverters~\cite{nguyen2020electric}. In this work, FDIA refers to control logic manipulation. Cyber-physical attacks are becoming a tangible risk: five consequential attacks were reported in 2023~\cite{DOE}, and fifteen FDIAs have been noted over the past ten years~\cite{HABIB2023}. The FDIAs can lead to improper control decisions that compromise the stability and reliability of power grid operations.

To address the risk of FDIAs, research efforts have focused on constructing accurate detection methods. FDIA detection methods typically rely on the comparison of a reference state (without FDIAs) and the actual observed state. This approach can take the form of a state predictor~\cite{sahu2024} or the construction of a state embedding~\cite{Aboelwafa}. Discrepancies between reference and observed states are then used to decide whether an FDIA has occurred. To realize this detection principle, a data-based approach would involve training a model that distinguishes between FDIA data and benign data. This often requires some level of supervised training where one assumes how FDIAs occur~\cite{Ying,Habibi}. Given an offline and fixed dataset, the neural network (NN)-based detection method can achieve promising performance via supervised learning. Ref.~\cite{Dehbozorgi2025} proposes an ensemble-based model to address the highly imbalanced nature of FDIA datasets. However, the performance of the detection model is only tested against a fixed adversarial dataset and does not consider adaptive or evolving attacks.
If the decision boundary between FDIA data and benign data is imperfect, the FDIA detection may become vulnerable to impactful though stealthy adversarial examples \cite{liu2011false}. Ref.~\cite{selim5346040} builds an enhanced adversarial model to incorporate uncertain samples that reside near the class decision boundary. For anomaly detection in general, iterative adversarial generation of anomalous data was shown to outperform state-of-the-art detection methods, especially with small datasets \cite{do2025swift}. These proposed methods improve the detection method by generating a more comprehensive and challenging set of attack scenarios.

Constructing FDIAs that bypass widely adopted detection methods has been the object of several research efforts ~\cite{standen2023}. In the power system domain, constrained optimization~\cite{Jafari,Choraria} and adversarial learning~\cite{Jiwei} strategies have been used to defeat bad data detection models. Ref.~\cite{Jiwei2024} proposes a multi-label adversarial attack framework, which can be more challenging to detect than single-label adversarial attacks. Ref.~\cite{WANG2024} utilizes the topological features of the power grid to improve the detection capability and generalization. The same optimization problem can be formulated with reinforcement learning (RL), which can be used for arbitrarily complex systems while generating hitherto unseen FDIAs by interacting with an environment~\cite{Romesh}. This approach has been used to improve detection methods in the context of frequency control~\cite{Mohamed} and state estimation problems~\cite{HUANG2023, Xiaohong}. To improve data-based defenders, Ref.~\cite{aslami2024} showed that the generated adversarial data can be ingested with a continual learning approach: there, a defender and an adversary are periodically and sequentially updated. An important drawback of this approach is that it is prone to catastrophic forgetting. These previous works only utilized a single RL agent to serve as either adversary or defender, but did not consider concurrent adversarial training. By contrast, we propose a multi-agent reinforcement learning (MARL) framework for concurrently training an FDIA adversary and defender.

The MARL framework has been applied to complex problems in cyber-physical systems that require cooperation among agents \cite{Yarahmadi2023}. For example, Ref. \cite{Amullen2016} uses separate defender agents that detect FDIAs at each substation. In contrast, this work explores the MARL framework with \textit{competitive agents} interacting in a \textit{non-stationary environment}. To address this, the learning environment and reward function are strategically designed to achieve the simultaneous performance improvement of both the adversary and defender agents. Even if the adversary continuously generates novel FDIAs to disrupt the system, the defender can promptly adjust its behavior and update its defense strategy accordingly. 

However, training the MARL framework can be challenging because two interacting agents have opposite optimization objectives. Learning from scratch through competitive interactions in a complex physical environment requires extensive trial-and-error exploration. Transfer learning enables the RL agent to leverage prior knowledge from an expert model to accelerate the learning process. Ref.~\cite{xu2021} improves FDIA detection in real-world transmission grids that may lack line parameters by transfer learning from the source domain, simulated grids with abundant simulation data. In power system applications ~\cite{Xiangyu2023, kejun2025}, transfer learning often utilizes a warm-start strategy to help the NN escape local minima and improve the training convergence rate. Warm-starting the NN weights can dramatically accelerate training compared to fresh random initializations ~\cite{NEURIPS2020}. 

The contributions of this paper are that:
\begin{itemize}
    \item Without prior knowledge, a defender trained with the proposed MARL framework can handle continuously varying unknown adversarial attacks launched by an adversary.
    
    \item Prior knowledge from an offline-trained defender can be retained through a transfer learning approach.
    
    \item The proposed method can locate FDIA attacks in both transmission and distribution systems.

\end{itemize}

\textit{Notation}: Upper (lower) boldface letters denote matrices (column vectors). Sets are represented by calligraphic letters. $|\mathcal{S}|$ represents the size of the set $\mathcal{S}$. $(\cdot)^{\top}$ denotes the vector/matrix transpose. 

Sec.~\ref{sec:prob} formulates the FDIA detection problem. Sec.~\ref{sec:method} describes the MARL framework and the transfer learning strategy. Two FDIA and detection scenarios are investigated, voltage control for the distribution grid in Sec.~\ref{sec:vol_reg} and frequency regulation for the transmission grid in Sec.~\ref {sec:fre_control}. Sec.~\ref{sec:results} evaluates its performance. Finally, Sec. \ref{sec:conclusion} concludes the manuscript.

\section{Problem formulation} \label{sec:prob}
This section describes the generic problem formulation for the FDIA and the detection mechanism. The power system is time-constrained, and each episode consists of $T$ timesteps spanning the control time interval. Let $\mathcal{T} \coloneq \{0, \cdots T-1\}$ collect the time steps in the control horizon. In the absence of detection, the adversary aims to disrupt the nominal operating status of the power grid, which can be expressed as:
\begin{subequations}
    \begin{align}
        \max_{\tilde{u}_{i,t}, \mathcal{N}^{a}_t} \, \, & \sum_{t \in \mathcal{T}} J(\mathbf{y}_t, \mathbf{y}_t^{\text{ref}}) \label{eq:obj}\,, \\
        \textrm{s.t.} \quad
        & u_{i,t} = \begin{cases}
            \tilde{u}_{i,t}\, , & i \in \mathcal{N}^{a}_t, \\
            u_{i,t}\, , & \text{otherwise}.
        \end{cases} \label{eq:attack} \\
        &\mathbf{\dot{z}}_t = \mathbf{f}(\mathbf{z}_t, \mathbf{u}_t)\, , \label{eq:dynamic} \\
        & \mathbf{y}_t = \mathbf{g}(\mathbf{z}_t, \mathbf{u}_t) \, , \label{eq:stat}
    \end{align}
\end{subequations}
where $\mathbf{z}_t$ denotes the system states and $\mathbf{y}_t$ denotes the variables of interest, which serve as the main indicators for identifying the occurrence of an FDIA. The adversary launches the FDIA by tampering with the control parameters of the inverter. Let $\mathcal{N}^a_t$ collect the compromised inverters whose control parameters have been modified by the adversary at timestep $t$. $u_{i, t}$ and $\tilde{u}_{i,t}$ denote the reference and the tampered control parameters, respectively. As shown in Eq.~\eqref{eq:obj}, the adversary maximizes the difference between the reference system state and the abnormal system state under FDIA, where $J$ is the metric that quantifies the physical instability induced by the FDIA. Eq.~\eqref{eq:attack} indicates that the compromised inverter uses the manipulated control parameter, while all the other inverters use the reference control parameters. Eq.~\eqref{eq:dynamic} describes the dynamical constraints (denoted by $\mathbf{f}$) of the system, such as dynamic behaviors of the inverters in frequency and voltage control problems. Eq.~\eqref{eq:stat} describes the grid operational constraints (denoted by $\mathbf{g}$), such as the power flow balance equations and the operating limits of the inverter. 

The primary goal of the defender is to locate the inverter that is under attack. In practice, the defender is a multiclass data-based classifier. The defender makes decisions based on the observed system states, such as voltage magnitudes, phase angles, and grid frequency.  

\section{Proposed MARL framework with transfer learning}\label{sec:method}
\begin{figure}[ht!]
    \centering
    \includegraphics[width=1\linewidth]{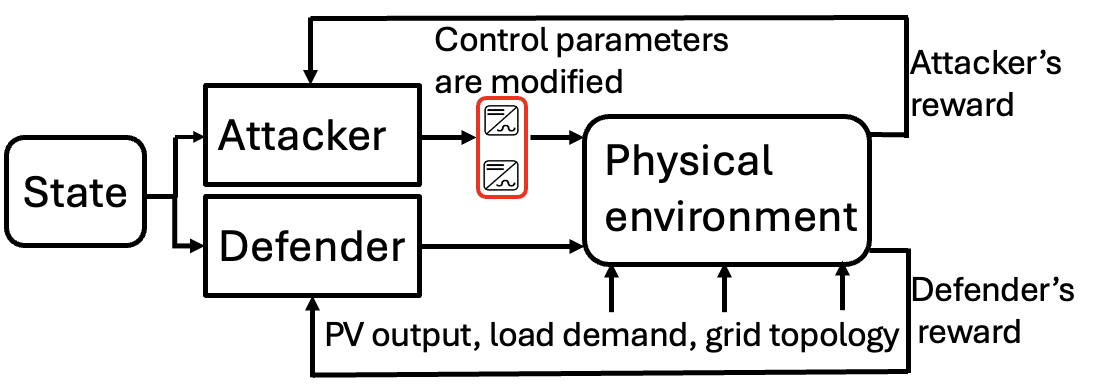}
    \caption{Multi-agent RL framework for the FDIA detection.}
    \label{fig:overall_architecture_all}
\end{figure}

We propose an MARL framework that jointly optimizes the adversary and defense strategies through interaction in a shared physical environment. MARL framework can utilize adversarial co-training between two agents, i.e., adversary and defender, to acquire defensive capabilities and improve overall system reliability \cite{pan2025}. Fig.\ref{fig:overall_architecture_all} depicts the proposed MARL framework for proactive FDIA detection. The adversary and defender aim to discover effective policies by competing against each other. They can adaptively learn and improve their strategies in response to the changing behavior of opponents. The adversary generates FDIAs that are both stealthy and impactful, and the detector learns to defend against stealthy and impactful attacks absent from its original dataset. The concurrent training allows to 1) focus the defender updates only on impactful FDIA and 2) expose both the defender and attacker to diverse and ever-changing environments. After the training completes, the defender is expected to identify the stealthy but impactful attacks.

\subsection{MARL framework}
The defender agent aims at detecting the bus under attack. The adversary agent seeks to disrupt the power grid's nominal operating states while avoiding being captured by the defender. The objective formulations are detailed as follows.  

\subsubsection{Defender}
The defender agent serves as a multiclass classifier and aims to identify the attack status, including the location of the attacked bus index or the identification that no attack has occurred. If the defender successfully identifies the attack status, it receives a reward $r > 0$ and a penalty $-r$ otherwise. The objective function of the defender can be expressed as:
\begin{subequations}
\begin{align}
\max \, \, \sum_{t \in \mathcal{T}_d}  \left( -(1 - D_{t}^{\text{suc}}) r + D_{t}^{\text{suc}} r \right ),  \label{eq:r_d_all} \\
D_{t}^{\text{suc}} = 
\begin{cases}
1 \, , & \text{Captured by the defender at $t \in  \mathcal{T}_d$}, \\
0 \, , &  \text{Escape from detection  at $t \in  \mathcal{T}_d$}, 
\\
0 \, , & \text{Non-detection time $t \in  \mathcal{T} \setminus \mathcal{T}_d$} \, ,
\label{eq:r_a_all}
\end{cases} 
\end{align}
\end{subequations}
where $\mathcal{T}_d$ denotes the detection time at which the defender tries to detect malicious activity.

\subsubsection{Adversary}
On top of the physical network, there exists a cyber-network that allows for data exchange between smart devices (including inverters) and the main control center. There are multiple ways for an adversary to compromise the cyber network~\cite{Mehmetcan2021}. Here, it is assumed that an adversary has already compromised and gained access to all the inverters connected to the cyber network. The scenario modeled in the proposed framework is one where the adversary decides to conduct an FDIA by modifying the inverter's control parameters, attempting to induce as much disturbance to the physical system as possible while avoiding detection. The objective of the adversary is:
\begin{subequations}
    \begin{align}
        \max_{\tilde{u}_{i,t}, \mathcal{N}^{a}_t} \, \, & \sum_{t \in \mathcal{T}} \left( (1 - D_{t}^{\text{suc}}) J(\mathbf{y}_t, \mathbf{y}_t^{\text{ref}}) + D_{t}^{\text{suc}} p  \right )\,
    \end{align}
\end{subequations}
where $p < 0$ is a penalty value given to the adversary when detected by the defender at the detection time. A large reward value reflects that the adversary induced system disruptions and that it escaped the defender.

\subsection{Transfer learning}
Ideally, the adversarial training of the defender refines an offline-trained one instead of training it from scratch. This way, expert knowledge can be distilled into the defender, and it can still be made robust to adversarial examples. To achieve this goal, we use the offline defender to initialize the policy neural network of the MARL-D. The goal of this procedure is to improve the defense strategy of the MARL-D against new adversarial attacks, while retaining the prior knowledge of the offline-trained defender. The training of MARL with transfer learning involves two stages, including pre-training and fine-tuning. In the pre-training stage, the expert model is trained based on abundant synthetic benign and abnormal data. Then, the MARL defender starts with a promising defense policy against certain attacks, but continues to be fine-tuned through online interactions with diverse attacks generated by the adversary.

In the remainder of the text, MARL-D refers to an agent trained from a random weight initialization, while a TF-MARL-D is a defender for which transfer learning is used, and TF-MARL-A denotes its corresponding adversary. Note that transfer learning is only applied to the defender; MARL-A and TF-MARL-A always use a random weight initialization. This is because while the defender is expected to retain knowledge to defend against FDIAs in general, no baseline performance is expected from the adversary.

The offline defender plays an essential role in improving the performance of the TF-MARL-D due to the knowledge transfer learning process. The offline defender is trained in a supervised fashion and falls into the multiclass classification task. Its performance can be sensitive to how balanced the training dataset is across classes under different attack scenarios. The dataset construction is described hereafter. At timestep $t$, if an FDIA occurs, the adversary modifies the controllable parameters of inverter $i$. Otherwise, the unaltered controllable coefficients are used. Throughout the episode, the training data is constructed by randomly selecting timesteps where an FDIA occurs. Let $T_a \in (0, 1)$ denote the fraction of episode steps where an FDIA occurs, and different setup values simulate different adversarial scenarios (e.g., $T_a = 0.8$ indicates an FDIA occurs for $80\%$ of the episode). The offline defender will serve as a baseline for comparison with the MARL defender. 

Ref.~\cite{sahu2024} justifies that the defender does not require inference at every timestep, thereby reducing the computational cost of each RL episode. Specifically, a detection window is introduced to reduce the detection frequency. Although the detection is called periodically every $d$ timesteps, it can detect FDIAs that occurred over the past $d-1$ timesteps~\cite{sahu2024}. The ensemble of detection timesteps is denoted by $\mathcal{T}_d$ and $\operatorname{card}(\mathcal{T}_d) = T/d$.
\begin{figure}[ht!]
    \centering
    \includegraphics[width=0.9\linewidth]{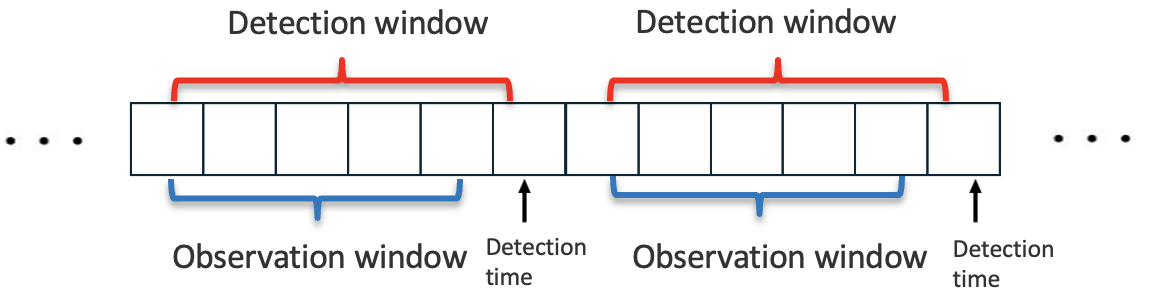}
    \caption{Schematic illustration of the detection scheme.}
    \label{fig:detection_time}
\end{figure}

\section{Case study I: Voltage regulation in the distribution grid}\label{sec:vol_reg}
\subsection{Voltage regulation problem formulation}
Consider a tree-structured distribution grid consisting of $N$ buses. The single-phase load can be connected to the three-phase network, which may lead to an unbalanced system. Let $\mathbf{p}_{d} \in \mathbb{R}^{N_d}$ and $\mathbf{q}_d\in \mathbb{R}^{N_d}$ collect the active and reactive load demands of all the loads. Each load is assumed to be connected with an inverter. Let $\mathbf{p}_{g} \in \mathbb{R}^{N_g}$ and $\mathbf{q}_{g} \in \mathbb{R}^{N_g}$ collect the corresponding power generation outputs from their associated inverters. Let $\mathbf{V}^{ph} \in \mathbb{R}^{3N} $ collect the voltage magnitudes of all the buses. The distflow equations $\mathbf{F}(\cdot)$ can be expressed as:
\begin{equation}
    \mathbf{V}^{ph} = \mathbf{F}(\mathbf{p}_d - \mathbf{p}_g, \mathbf{q}_d - \mathbf{q}_g)\, . 
\end{equation}

\begin{figure}[ht!]
    \centering
    \includegraphics[width=1\linewidth]{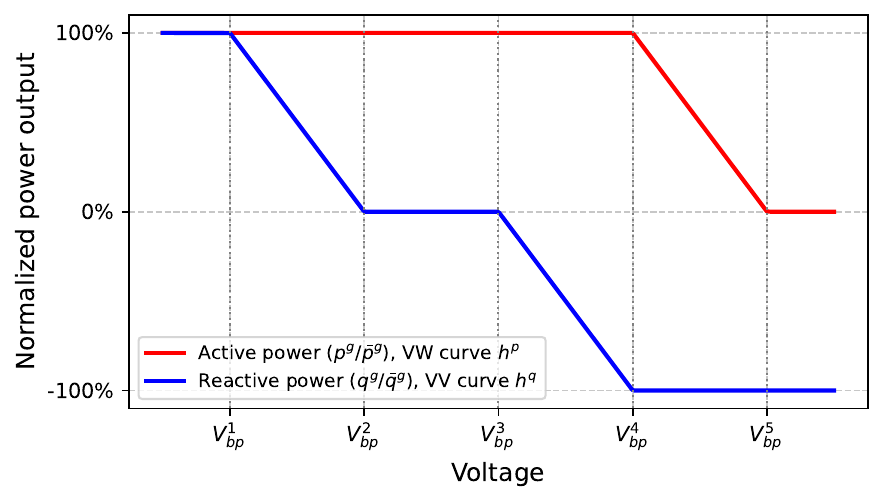}
    \caption{Illustration scheme of the VW and VV control functions.}
    \label{fig:vvvw}
\end{figure}

\begin{figure}[ht!]
    \centering
    \includegraphics[width=1\linewidth]{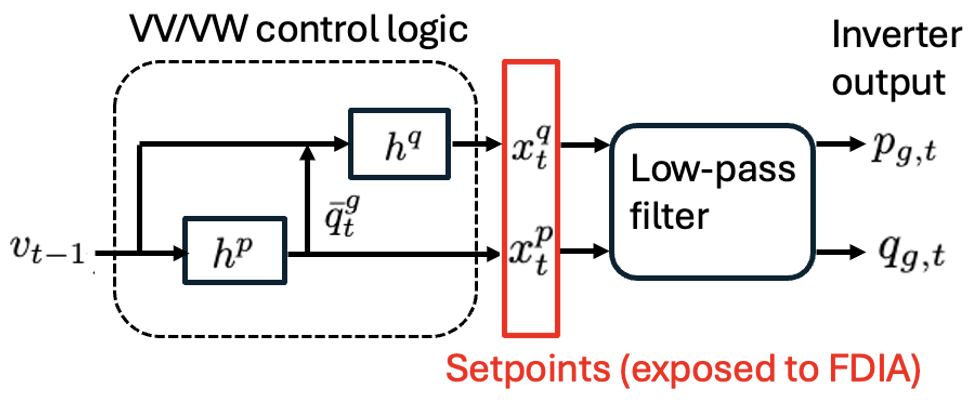}
    \caption{The structure of the inverter and the FDIA scheme.}
    \label{fig:vvvw_attack}
\end{figure}

We assume the inverters are operated following the Volt-VAR (VV) and Volt-Watt (VW) control logic \cite{Roberts2021}. To regulate the voltage, the control logic enables the inverter to dynamically adjust the active and reactive power output in response to the changing local conditions. The scheme of VV/VW control logic is depicted in Fig. \ref{fig:vvvw}. The voltage breakpoints of the inverter $V_{\text{bp}}^i$, $i\in\{1, 2, 3, 4, 5\}$ are assumed to be given and fixed. Let $x_t^p$ and $x_t^q$ denote the active and reactive power setpoints, respectively, and they can be calculated by \cite{Roberts2021}:
\begin{subequations}
  \begin{align}
    x_t^p :&= h^p(v_{t-1}) \,, \text{where}\, x_t^p \in [0, \bar{p}_{t}^{g}] \\
    x_t^q :&= h^q(v_{t-1}) \,, \text{where}\, x_t^q \in [-\bar{q}_{t}^{g}, \bar{q}_{t}^{g}]
\end{align}  
\end{subequations}
where $h^p(\cdot)$ and $h^q(\cdot)$ denote the VW and VV control functions, respectively. $\bar{p}_{t}^{g}$ denotes the maximum active power generation output of the inverter at time $t$, which depends on the available solar irradiation. The maximum absolute reactive power output of the inverter can be obtained as:
\begin{equation}
     \bar{q}_{t}^{g} = \sqrt{(s_g^{\text{max}})^2 - (x_t^p)^2},
\end{equation}
where $s_g^{\text{max}}$ represents the maximum capacity of the inverter. Instead of directly applying the instantaneous setpoints suggested by the VV/VW control logic, a first-order low-pass filter is embedded within the inverter to avoid rapid changes and yield smoother output \cite{Ciaran2020}. The power outputs of the inverter are given as:
\begin{subequations}
  \begin{align}
    p^g_t &= (1-\alpha) p^g_{t-1} + \alpha x_t^p \,, \\
    q^g_t &= (1-\alpha) q^g_{t-1} + \alpha x_{t}^q \,,
\end{align}  
\end{subequations}
where $p^g_t$ and $q^g_t$ denote the active and reactive power outputs of the inverter, respectively. $\alpha \in [0, 1]$ denotes the filter time constant, which serves as a smooth parameter. Figure~\ref{fig:vvvw_attack} demonstrates the structure of the inverter, and the FDIA attack considered in the study is that the adversary can overwrite the setpoints of the inverters \cite{Hossen2020}. 

The objective of the adversary is to induce voltage instability by modifying the setpoints of the inverter, making the VV/VW control logic invalid. The adversary can compromise one inverter at a time. In the absence of detection, the adversary aims to solve the following optimization problem:
\begin{subequations}
 \begin{align}
    \max_{\tilde{x}_{i, t}^p, \tilde{x}_{i, t}^q, \mathcal{N}_{t}^{{a}}} \, \, & \sum_{t \in \mathcal{T}} \sum_{i \in \mathcal{N}_d}(v_{i, t} - 1)^2 \, ,  \label{eq:ob1} \\
    \textrm{s.t.} \quad 0 &\leq \tilde{x}_{i, t}^p \leq \bar{p}_{i, t}^{g} \, , i \in \mathcal{N}_t^{a} \\ 
     -\bar{q}_{i, t}^{g} &\leq \tilde{x}_{i, t}^q \leq \bar{q}_{i, t}^{g} \,, i \in \mathcal{N}_t^{a} \\
    \tilde{p}^g_{i, t} &= (1-\alpha) p^g_{i, t-1} + \alpha \tilde{x}_{i, t}^p \,,  i \in \mathcal{N}_t^{a} \label{eq:low_p} \\
    \tilde{q}^g_{i, t} &= (1-\alpha) q^g_{i, t-1} + \alpha \tilde{x}_{i, t}^q \,, i \in \mathcal{N}_t^{a} \\
    |\mathcal{N}_{t}^{a}| & = 1, \,  \mathcal{N}_{ t}^{a} \subseteq \mathcal{N}_g \label{eq:change_vol} \\ 
    p^g_{i, t}= \tilde{p}^g_{i, t} , \,& q^g_{i, t} = \tilde{q}^g_{i, t}\, ,i \in \mathcal{N}_t^{a}\, ,  \label{eq:attackbus_vol} \\
    p^g_{i, t} = p^g_{i, t} , \, & q^g_{i, t} = q^g_{i, t}\, ,i \in \mathcal{N}_g \setminus \mathcal{N}_t^{a} \, , \label{eq:normalbus_vol} \\
    \mathbf{V}^{ph}_{t} &= \mathbf{F}(\mathbf{p}^d_t - \mathbf{p}^g_t, \mathbf{q}^d_t - \mathbf{q}^g_t) \, , \label{eq:dist} 
\end{align}   
\end{subequations}
Eq.~\eqref{eq:change_vol} implies that at most one inverter is under attack by the adversary at each timestep. Eqs.~\eqref{eq:attackbus_vol} and ~\eqref{eq:normalbus_vol} show only that the attacked inverter uses modified power setpoints. The objective function \eqref{eq:ob1} indicates that the adversary tamper with active and reactive power setpoints of the $i$-th inverter (denoted by $\tilde{x}_{i, t}^p$ and $\tilde{x}_{i, t}^q$) to maximize the voltage deviation in the control horizon. 

\subsection{MARL framework for FDIA detection in voltage control}
\begin{figure}[ht!]
    \centering
    \includegraphics[width=1\linewidth]{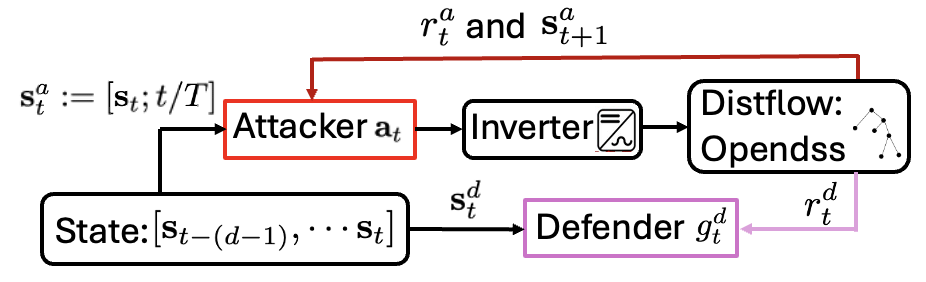}
    \caption{Multi-agent framework for the FDIA detection of voltage control in the distribution grid.}
    \label{fig:overall_architecture_vol}
\end{figure}

Fig.~\ref{fig:overall_architecture_vol} illustrates the proposed MARL framework for FDIA detection in voltage control. 
\subsubsection{Adversary}
The adversary's observation space, $\mathbf{s}_t^a$, is the concatenation of the power system operating states $\mathbf{s}_t = [\mathbf{p}_d^\top, \mathbf{q}_d^\top, \mathbf{p}_g^\top, \mathbf{q}_g^\top, \mathbf{V}^\top]^\top \in \mathbb{R}^{3N_d + 2N_g}$ and the normalized simulation step $\frac{t}{T}$. The normalized simulation step provides the current time to the adversary. The adversary's action space is $\mathbf{a}_t = [g_t^A, c_t^p, c_t^q, m_t]$, where $g_t^A \in \{-1, 0, \cdots, N_g-1 \}$ denotes the index of the attacked inverter or $-1$ for no attack. To respect the apparent power capacity of the inverter, we introduce the bounded $c_t^p \in [0, 1]$ and $c_t^q \in [-1, 1]$ to formulate the modified active and reactive power setpoints:
\begin{subequations}
   \begin{align}
    \tilde{x}_{t}^p :&= c_t^p \bar{p}_{t}^{g} \, ,  \\
    \tilde{x}_{t}^q :&= c_t^q \bar{q}_{t}^{g} .
   \end{align} 
   \label{eq:action_vol}
\end{subequations}
Given the detection time $t \in \mathcal{T}_d$, the detection window is defined as $\mathcal{D}_{t}\coloneq [t-(d-1), \cdots, t]$. We define $g^a$ to record the attack bus index over $\mathcal{D}_{t}$ and initialize it as $-1$. To ensure that at most one bus is under attack over the detection window $\mathcal{D}_{t}$, a boolean variable $m_t$ is introduced to mute the invalid attack index $g_t^A$. For example, if two different inverters are selected over $\mathcal{D}_{t}$, $m_t$ mutes the latter choice and sticks to the former. The post-processing action mask mechanism can be expressed as:
\begin{align}
g^a_t = 
\begin{cases}
g^A_t \, , & \text{if} \, g^a = -1\, , \\ 
g^a\, , &  \text{if} \,  m_t = 0 \land g^a \neq -1\, , \\
-1 \, , &  \text{if} \,  m_t = 1 \land g^a \neq -1,
\end{cases}   
\label{action_pp}
\end{align}
\noindent where, $\land$ denotes the logical AND operator. Once the bus to attack is chosen with $g_t^A$, the binary variable $m_t$ is used to decide whether to attack ($m_t = 0$) or mute the attack ($m_t = 1$). 

The stealthiness constraint is included in the new objective function Eq.~\eqref{eq:problem_dec_vol} through a penalty $p$ for being detected.
\begin{subequations}
 \begin{align}
    \max_{\tilde{x}_t^p, \tilde{x}_t^q, \mathcal{N}_{t}^{a}} \, \, & \sum_{t \in \mathcal{T}} \left( (1 - D_{t}^{\text{suc}})  \sum_{i \in \mathcal{N}_d} (v_{i, t} - 1)^2 + D_{t}^{\text{suc}} p  \right ) \, ,  \label{eq:ob_RL_vol} \\
    \text{s.t.} \quad & \text{Eqs.} \, \eqref{eq:low_p} - \eqref{eq:dist}\, , \eqref{eq:action_vol} \, .
\end{align}   
\label{eq:problem_dec_vol}
\end{subequations}
Based on the objective~\eqref{eq:ob_RL_vol}, the instantaneous reward at time $t$ is designed as:
\begin{align}
r_t^a = 
\begin{cases}
p \, , & g_t^a = g_t^d \, \,  \land \, \, g_t^a \neq -1 \, , t \in  \mathcal{T}_d \\
c_v\frac{\sum_{i \in \mathcal{N}_d} (v_{i, t} - 1)^2}{|\mathcal{N}_d|} \, , &  \text{otherwise} \, .
\end{cases} 
\end{align}
where $c_r$ denotes a scale factor.

\subsubsection{Defender}
The defender’s observation space is $\mathbf{s}_t$, and the action space is $g^d_t \in \{-1, 0, \cdots, N-1 \}$. The defender aims to locate the attacked bus index or identify no attack. Based on the objective function of the defender \eqref{eq:r_d_all}, the defender maximizes the cumulative reward in the control horizon:
\begin{equation}
    \max_{g_t^d} \, \, \sum_{t \in \mathcal{T}} r_{t}^{d}
\end{equation}
where the instantaneous reward at time $t$ is defined as:
\begin{align}
r_t^d = 
\begin{cases}
r \, , & g_t^d = g_t^a \,, t \in  \mathcal{T}_d \, ,\\
-r \, , & g_t^d \neq g_t^a \, , t \in  \mathcal{T}_d \,, \\
0 \, , &  t \in  \mathcal{T} \setminus \mathcal{T}_d \, .
\end{cases} 
\label{eq:reward_d}
\end{align}

\section{Case study II: Frequency control in the transmission grid} \label{sec:fre_control}
\subsection{Frequency control problem formulation}
We consider a lossless power flow model in the transmission grid. In the primary control problem, we assume the reactive powers are ignored, and the voltage magnitudes are well-regulated. The environment is chosen to keep the computational cost of the RL training loop low while still approximating the frequency dynamics. The system state $\mathbf{s} \coloneq [\boldsymbol{\theta}^\top ; \boldsymbol{\omega} ^\top]^\top \in \mathbb{R}^{2N}$ contains phase angles and frequency deviation of all buses. The primary frequency dynamics can be expressed as the swing equation \cite{Wenqi}:
\begin{subequations}  \label{eq:swing}
\begin{align}
\dot{\theta_i} &= \omega_i\, , \\
M_i\dot{\omega_i} &= p_{i} - p^ {\text{IBR}}_i(\omega_i) - D_i \omega_i - \sum_{j=1}^N B_{ij} \sin{(\theta_i - \theta_j)}  \, , 
\end{align}
\end{subequations}
where $M_i$ and $D_i$ represent the inertia and damping coefficients of the $i$-th bus, respectively. $p_i$ denotes the net power injection of bus $i$ and $B_{ij}$ is the ($i, j$)-th element in the susceptance matrix. $p^{\text{IBR}}_i$ represents the active power output from the inverter-based resources at bus $i$, which can be controlled by the linear droop controller, i.e., $p^ {\text{IBR}}_i\coloneq k_i^{\text{ref}} \omega_i$. The droop coefficients can be designed using conventional optimization methods to stabilize the primary frequency under a random off-equilibrium initial condition \cite{Wenqi}. 

In the absence of detection, the adversary aims to solve the following optimization problem \cite{Romesh}:
\begin{subequations}
 \begin{align}
    \max_{\tilde{k}_{i, t}} \, \, & \sum_{t \in \mathcal{T}} \sum_{i \in \mathcal{N}}|\omega_{i, t}| - |\omega_{i, t}^{\text{ref}}| \, ,  \label{eq:ob1_fre} \\
    \text{s.t.} \quad & \text{Eq.} \, \eqref{eq:swing} \, , \label{eq:swing1}  \\
    & p^{\text{IBR}}_{i,t} = \tilde{k}_{i, t} w_{i, t} \, , \label{eq:ibr}  \\
    & \|\mathbf{k}^\text{ref} - \tilde{\mathbf{k}}_{t} \|_0 \leq 1 \, ,  \label{eq:change}
\end{align}   
\end{subequations}
where $\omega_i^{\text{ref}}$ denotes the frequency deviation of bus $i$ using the unaltered droop controller. The objective function \eqref{eq:ob1_fre} indicates the adversary aims to tamper with the droop coefficient (denoted by $\tilde{k}_i$) to maximize frequency deviation in the control horizon. Eq.~\eqref{eq:change} ensures that at most one droop coefficient can be modified by the adversary at any timestep, where $\tilde{\mathbf{k}}_t \in \mathbb{R}^N$ denotes the vector of (tampered) droop coefficients at time $t$ and $\mathbf{k}^\text{ref}$ are the unaltered droop coefficients of all buses. 

\subsection{MARL framework for FDIA detection in frequency control}
\begin{figure}[ht!]
    \centering
    \includegraphics[width=1\linewidth]{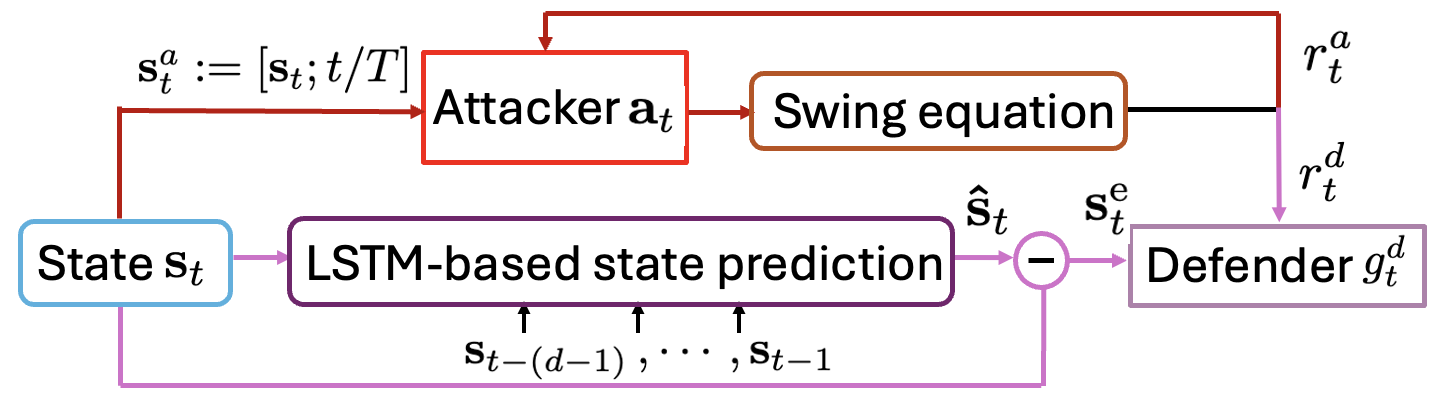}
    \caption{Multi-agent framework for the FDIA detection in frequency control.}
    \label{fig:overall_architecture}
\end{figure}

Fig.~\ref{fig:overall_architecture} illustrates the proposed MARL framework for proactive FDIA detection in frequency control. The LSTM-aided defender aims to identify the attacked inverter. The adversary agent seeks to disrupt the power grid frequency while avoiding being captured by the defender. 

\subsection{Adversary}
The adversary's observation space is the concatenation of the system states and the normalized simulation step, i.e. $\mathbf{s}_t^a \coloneq[\mathbf{\boldsymbol{\theta}}^\top, \boldsymbol{\omega}^\top, t/T]^\top$. The adversary's action space is $\mathbf{a}_t = [g_t^A, c_t, m_t]$, where $c_t$ represents the value with which the droop coefficient is replaced ($\tilde{k}_{i, t} = c_t$). For simplification and consistency with Refs.~\cite{sahu2024,Romesh}, $c_t \in \{-1, \, 0, \, 1\}$. The new objective function with the stealthiness constraint can be expressed as:
\begin{subequations}
 \begin{align}
    \max_{\tilde{k}_{i, t}} \, \, & \sum_{t \in \mathcal{T}} \left( (1 - D_{t}^{\text{suc}})  \sum_{i \in \mathcal{N}} (|\omega_i| - |\omega_i^{\text{ref}}| ) + D_{t}^{\text{suc}} p  \right ) \, ,  \label{eq:ob} \\
    \text{s.t.} \quad & \text{Eq.} \, \eqref{eq:swing}, \eqref{eq:ibr} - \eqref{eq:change} \, .
\end{align}   
\label{eq:problem_dec}
\end{subequations}
The instantaneous reward at time $t$ is designed as:
\begin{align}
r_t^a = 
\begin{cases}
p \, , & g_t^a = g_t^d \, \,  \land \, \, g_t^a \neq -1 \, , t \in  \mathcal{T}_d \\
r_t^\omega\, , &  \text{otherwise} \, , 
\end{cases} 
\end{align}
where $r_t^\omega \coloneq c_s \sum_{i \in \mathcal{N}}(|\omega_{i, t} | - |\omega_{i, t}^{\text{ref}} |) $ represents the scaled frequency deviation difference and $c_s$ is the scale parameter.

\subsection{LSTM-aided defender}
\begin{figure}[ht!]
    \centering
    \includegraphics[width=0.9\linewidth]{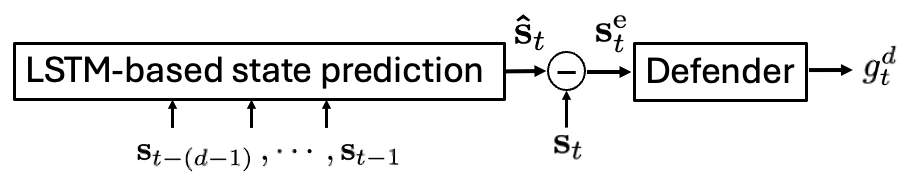}
    \caption{Schematic of the offline FDIA defender.}
    \label{fig:offline_dec}
\end{figure}

The choice behind an LSTM-based detection model is thoroughly discussed in Ref.~\cite{sahu2024}, and the instantaneous reward is detailed in Eq.\eqref{eq:reward_d}. The architecture of the offline defender is schematically shown in Fig.~\ref{fig:offline_dec}. The LSTM predicts the reference state at time $t$ based on the observed states over the past $d-1$ timesteps, i.e., $[\mathbf{s}_{t-(d - 1)}, \cdots \mathbf{s}_{t - 1}] \mapsto \mathbf{\hat{s}}_t$. The LSTM is held frozen during the MARL training process. The defender is a multiclass classifier that locates the attacked bus index based on the state prediction error, i.e., $\mathbf{s}^e_t \mapsto g^d_t $, where $\mathbf{s}_t^\mathrm{e}\coloneq c_{\text{wup}}(\mathbf{s}_t - \mathbf{\hat{s}}_t)$, where the class $-1$ implies that no attack occurs. 

Ref.~\cite{NEURIPS2020} points out that the warm-up strategy can improve training efficiency while sacrificing generalization performance on the testing samples in certain tasks. They propose some experimental techniques to mitigate this issue, such as shrinking the old weights and then adding random noise. In the experiments, we investigate how the generalization ability of warm-starting is affected by the distribution of the initial weights of the offline defender. To address this, a coefficient parameter $c_{\text{wup}}$ is adopted to scale the input features, which can shrink the NN weights to ensure a proper distribution and needs to be fine-tuned. 

\section{Numerical results} \label{sec:results}
The effectiveness of the proposed MARL framework on voltage control and frequency regulation is verified on the IEEE-37 bus system and the IEEE New England 39-bus system, respectively. We use the Ray library with TensorFlow to train MARL using the independent proximal policy optimization (PPO) algorithm. The training of competitive agents can be challenging because the reward of the adversary agent is dependent on the defender's performance. Specifically, the adversary agent will receive a penalty for being captured by the defender. Hence, the value of the penalty parameter $p$ will affect the adversary agent's action policy and needs to be fine-tuned. The chosen criterion is that the adversary agent introduces significant disruptions to the system, while the defender's detection accuracy is promising. We implement the simulations using the high-performance computing (HPC) system in parallel across 104 CPU cores. Sec.\ref{sec:vol_control_simu} and Sec.\ref{sec:fre_control_simu} show the numerical results of the voltage control and frequency regulation problems, respectively. 

\subsection{Voltage control} \label{sec:vol_control_simu}
\subsubsection{Simulation setup}
The IEEE-37 bus system is a three-phase unbalanced distribution grid, and the topology can be found in Ref.\cite{TAHERI2019}. The load demand data is modified based on real-time Portuguese electricity consumption, and the PV data is collected from a Belgian power network operator ~\cite{NEURIPS2021}. There are 30 loads and 30 inverters connected to the grid, and the line parameters are given by the OpenDSS \cite{dugan2012}. The dist flow equations are solved by the OpenDSS solver \cite{dugan2012}. We consider a control horizon of $5$ hours (10:00 AM–3:00 PM) in May. Given a 3-minute control interval, the total number of control steps per episode is $T=100$. The window length is set to $d = 6$, and the detection occurs every 18 minutes. The smooth parameter $\alpha$ is set to 0.3, and the breakpoints are $V_{\text{bp}} = \{0.95, 0.98, 1, 1.02, 1.05\}$. The reward scale factor $c_r$ is set to 100. The reward for a successful defense and the penalty for being captured are given as $r = -p = 0.01$. The entropy coefficient and the clip range for the PPO are set to $0.001$ and $0.2$, respectively. The learning rate is $5\times10^{-5}$. The policy neural networks of the adversary and defender contain two hidden layers, each consisting of $256$ neurons with a ReLU activation function. The rollout fragment length is set to $100$. 

\subsubsection{Detection comparison results}

\begin{figure}[h!]
    \centering
    \includegraphics[width=1\linewidth]{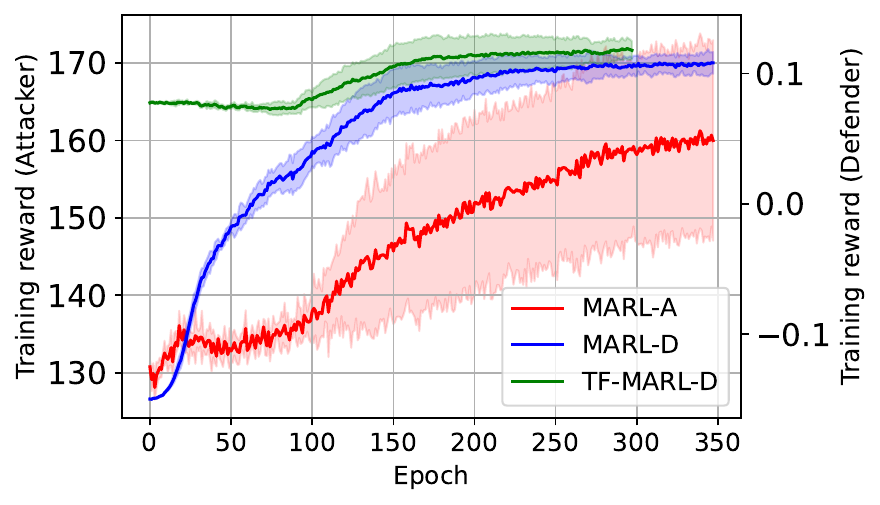}
    \caption{History of the agent's rewards as a function of epoch in the voltage control case study. The solid line and the shaded area are the mean and standard deviation of the reward obtained with five independent training runs (faded color curves).}
    \label{fig:train2}
\end{figure}

As shown in Fig.~\ref{fig:train2}, TF-MARL-D achieves a higher reward value than MARL-D, which verifies the effectiveness of the warm-up strategy. Thanks to the knowledge transfer process, TF-MARL-D retains the expert knowledge from the offline defender and obtains a much higher reward than MARL-D in the initial training stage. Hence, it will be beneficial to adopt the warm-up strategy when given limited training time and computational resources.

Table~\ref{tab:att_def_acc} shows the detection accuracy obtained with the MARL-D, TF-MARL-D, and Offline-D against all their adversary counterparts. The poor performance of MARL-D against Offline-A is an important result that shows the limits of the MARL framework proposed here: although the diversity of FDIA and defender is higher than without a concurrent training approach \cite{aslami2024}, a defender trained only with the MARL framework cannot be expected to be capable against all possible FDIA. In particular, it performs poorly against synthetic FDIA. 
Compared to MARL-D, TF-MARL-D doubles the detection accuracy against offline-A and appears as a reasonable cure to the problem discussed above. TF-MARL-D also achieves comparable performance to MARL-D against online MARL attacks, and better performance than Offline-D against the unseen attacks generated by MARL-A. As can be expected, Offline-D achieves the best detection accuracy against Offline-A. However, it achieves the worst detection accuracy against adversarial attacks generated by TF-MARL-A and MARL-A.

\begin{table}[h!]
\centering
\caption{Accuracy (\%) of each adversary against each defender in the voltage control case study.}
\label{tab:att_def_acc}
\setlength{\tabcolsep}{10pt}
\renewcommand{\arraystretch}{1.1}
\begin{tabular}{lccc}
\toprule
\textbf{Adversary} & \textbf{MARL-D} & \textbf{TF-MARL-D}  & \textbf{Offline-D}  \\
\midrule
MARL-A    &  91\%  &  86\% & 83\% \\
TF-MARL-A &  90\%  &  93\%  &  83\%\\
Offline-A &  42\% &  85\%  &  93\% \\
\hline
Mean & 78\%  & 88\% & 86\% \\
\bottomrule
\end{tabular}
\end{table}

As highlighted in Ref.~\cite{aslami2024} one of the challenges of continual reinforcement learning is catastrophic forgetting, i.e., the defender drastically forgets the previously learned policy upon learning new attacks \cite{aslami2024}. In the present case, TF-MARL-D still achieves $85\%$ detection accuracy against Offline-A which suggest that transfer learning can adress catastrophic forgetting (which was not the case in Ref.~\cite{aslami2024}). The amount of forgetting can also be evaluated with the adversary and defender checkpoints: if a defender performs well against the most recent adversary but not older adversary, it can point to gradual forgetting throughout training, in which case the defender performance might oscillate indefinitely. Table~\ref{tab:p_com_nowup_vol} shows the accuracy of the final trained defender's policy exercised against intermediate adversaries. It is clear that TF-MARL-D retains more information than MARL-D throughout training and that catastrophic forgetting is not significant.

\begin{table}[h!]
\begin{center}
\renewcommand{\arraystretch}{1.1}
\caption{Detection accuracy of MARL-D$_\text{(E280)}$ and TF-MARL-D$_\text{(E280)}$ against attacks generated by MARL-A and TF-MARL-A during training, respectively. The subscript denotes the corresponding policy at epoch 280.}
\label{tab:p_com_nowup_vol}
\begin{tabular}{ |c|c|c|c|c|c|c|c|c|c|c } 
\hline
MARL-A$_{(\cdot)}$ & E280 & E200 & E160 & E120 & E80 & E40
\\ \hline 
MARL-D$_\text{(E280)}$ & 92\%  & 94\% & 86\% & 68\% & 83\% & 89\% \\
\hline
\hline
TF-MARL-A$_{(\cdot)}$ & E280 & E200 & E160 & E120 & E80 & E40
\\ \hline 
TF-MARL-D$_\text{(E280)}$ & 94\% & 96\% & 96\% & 95\% & 90\% & 95\% \\
\hline 
\end{tabular}
\end{center}
\end{table}

\begin{figure*}[ht!]
    \centering
    \includegraphics[width=1\linewidth]{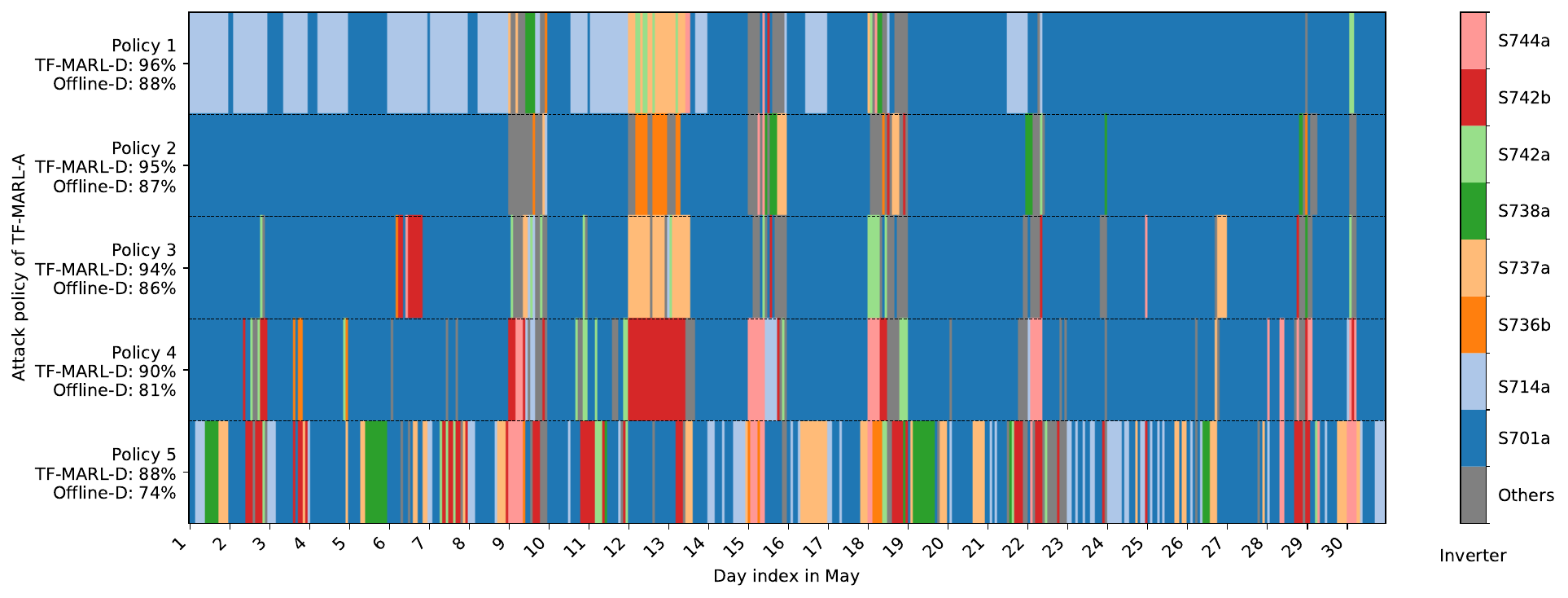}
    \caption{The attack strategies are generated by five independently trained TF-MARL-A agents in the voltage control case study. The y-axis shows the corresponding detection accuracy of the offline-D and TF-MARL-D defenders under the attack scenarios.}
    \label{fig:attack}
\end{figure*}

Fig.~\ref{fig:attack} shows the primary attacked inverter index each day and the detection accuracy. The inverter S701a is close to the substation (root) of the distribution grid, and it is likely to be attacked. The adversary may choose to attack different buses in one month because the load and PV data vary daily. Compared to other trained policies, Policy 5 adopts more varied actions, which implies the corresponding TF-MARL-A frequently changes the attacked inverter target. This makes detection more challenging than the deterministic attacks, which leads to the performance degradation for TF-MARL-D and Offline-D. However, TF-MARL-D still achieves 88\% detection accuracy, but Offline-D detection accuracy decreases to 74\%.

\subsection{Frequency regulation} \label{sec:fre_control_simu}
\subsubsection{Simulation setup}
The proposed framework is tested on the 10-bus Kron reduced IEEE New England 39-bus system, for which unaltered droop coefficients are provided in \cite{Wenqi}. Here, the control horizon is $0.05\rm{s}$ and the simulation interval is $0.01\rm{s}$, leading to a total number of steps per episode $T=500$. Consistently with Ref.~\cite{sahu2024}, the window length is set to $d = 6$ and the ensemble of detection timesteps is therefore $\mathcal{T}_d : = \{ 6, 12, \cdots, 498\} $. Initial conditions are constructed by superimposing disturbances onto the equilibrium state. Disturbances are sampled from $\mathcal{U}(-0.2, 0.2)$ for both the phase and frequency. The same initial condition is used in MARL and offline defender training. In the adversary reward function, we set $c_s = 0.1$ to avoid large training reward values for stabilizing the training process. The reward for a successful defense and the penalty for being captured are given as $r = -p = 0.1$, and $c_{\text{wup}}$ is set to $100$. The entropy coefficient and the clip range are set to $0.01$ and $0.2$, respectively. The LSTM is the same as the one described in Ref.~\cite{sahu2024} and uses 100 units. The policy neural networks of the adversary and defender contain two hidden layers, each consisting of $256$ neurons with a hyperbolic tangent ($\tanh$) activation function. The learning rate is set to $10^{-4}$, and the rollout fragment length is set to $500$.

\subsubsection{MARL detection vs Offline-D}
\begin{figure}[ht!]
    \centering
    \subfloat[Droop coefficient $c_t$ value. \label{fig:coe}]{%
       \includegraphics[width=0.48\linewidth]{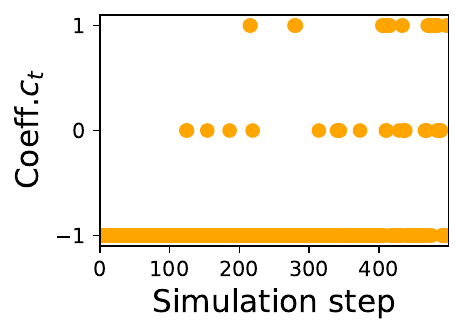}}
    \hfill\subfloat[Frequency deviation of each bus. \label{fig:fre_train}]{%
        \includegraphics[width=0.5\linewidth]{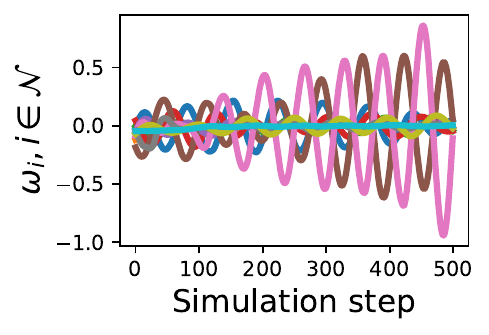}}
    \caption{MARL-A's action in one episode and the resulting frequency deviation.}
  \label{control_123_25p}
\end{figure}

In the following two figures, only the results of the training run that leads to the highest accuracy for the MARL-D are shown. As shown in Fig.~\ref{fig:coe}, the adversary typically chooses $c_t=-1$ because a negative value of $c_t$ leads to more impactful attacks \cite{Romesh}. Fig.~\ref{fig:fre_train} shows the resulting frequency instability of each bus caused by the MARL-A.

\begin{figure}[ht!]
    \centering
    \includegraphics[width=1\linewidth]{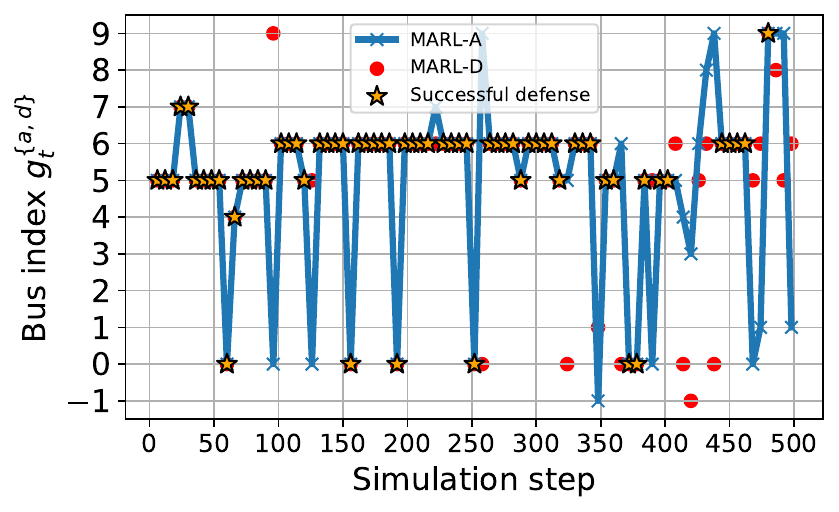}
    \caption{Attacked bus by MARL-A and the detected bus by MARL-D in the frequency regulation case study. A successful defense implies $g_t^a = g_t^d $ and the detection accuracy is $77\%$.}
    \label{fig:det_action}
\end{figure}

Ref.~\cite{wang2020locational} utilizes the convolutional neural network (CNN) as the classifier to detect the exact locations of FDIA based on the observed system states at each timestep. As shown in Table \ref{tab:p_com_nowup_fre}, the MARL-D also outperforms the offline, CNN-based defender against adversarial attacks in all the independent simulation runs. As expected, the MARL-D outperforms the offline detection when exercised against the MARL-A, but the accuracy gap between MARL-D and Offline-D is larger than in the distribution case. Fig.~\ref{fig:det_action} shows the attacked bus index and the identified attacked bus by the MARL-D based on the trained policy in the fifth run. Both the offline defender and the MARL-D successfully catch the attacked bus index $6$, which indicates that continuously attacking the same bus is likely to be captured. However, the MARL-D outperforms the offline defender when the adversary rapidly varies its action.


\begin{table}[h!]
\begin{center}
\renewcommand{\arraystretch}{1.1}
\caption{Detection accuracy (\%) under MARL-A's generated attacks of multiple independent training runs in the frequency regulation case study. The last column (to the right of the vertical divider) shows the average detection accuracy, and the same convention applies to the other tables. }
\label{tab:p_com_nowup_fre}
\begin{tabular}{ |c|c|c|c|c|c||c| } 
\hline
$\sum_{t \in \mathcal{T}}r_t^w$  & $44$  & $52$ & $34$ &  $52$ & $34$ & Mean
\\
\hline
MARL-D & 63\% & 65\% & 71\%  & 71\% & 77\% & \textbf{70}\% \\
\hline
Offline-D & 58\% & 60\% & 36\%  & 61\%  & 43\%  & 52\%\\ 

\hline
CNN-D \cite{wang2020locational} & 59\% & 63\% &  46\% & 59\% & 41\% & 54\% \\ 

\hline
\end{tabular}
\end{center}
\end{table}



\subsubsection{Warmed-up MARL detection} \label{sec:Warmed-up MARL detection}
In this section, we investigate whether it is reasonable to rely on an MARL training framework to construct a detection method that 
performs well on synthetic and adversarially generated attacks. 
For this experiment, the defender is exercised against synthetic FDIAs that would cause the largest system disruption, i.e., when the adversary chooses $c_t = -1$ over the entire control horizon (which is referred to as \textit{time-invariant} attack)~\cite{Romesh}.

\begin{figure}[ht!]
    \centering
    \includegraphics[width=1\linewidth]{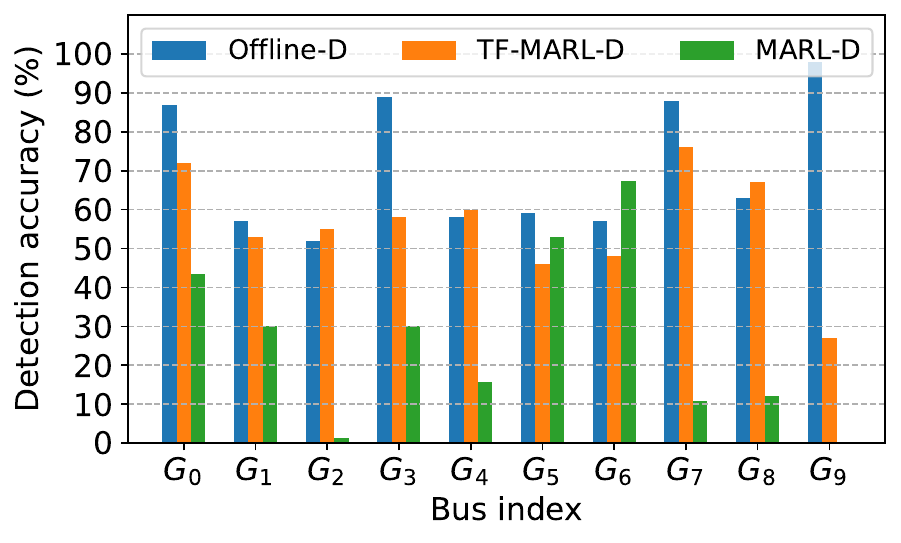}
    \caption{The detection accuracy under time-invariant attacks in the frequency regulation case study.}
    \label{fig:p_com_invi}
\end{figure}

Similar to the findings obtained in Case Study I on the distribution system, the offline defender outperforms the MARL-D except for bus $6$ (Fig.~\ref{fig:p_com_invi}), which is often targeted by the MARL-A (Fig.~\ref{fig:det_action}). This confirms that although MARL-A can discover a diverse set of attacks that help improve the detection performance of MARL-D, it cannot be expected to have exhaustively explored the space of FDIAs.
Just like in Case Study I, the transfer learning approach is a viable pathway to ensure that the defender is accurate on both the synthetic and the adversarially generated FDIA: Fig.~\ref{fig:p_com_invi} shows that the detection accuracy of TF-MARL-D are on par or even outperform (buses $2$, $4$, and $8$) the offline defender for time-invariant attacks. 

\begin{figure}[ht!]
    \centering
    \includegraphics[width=1\linewidth]{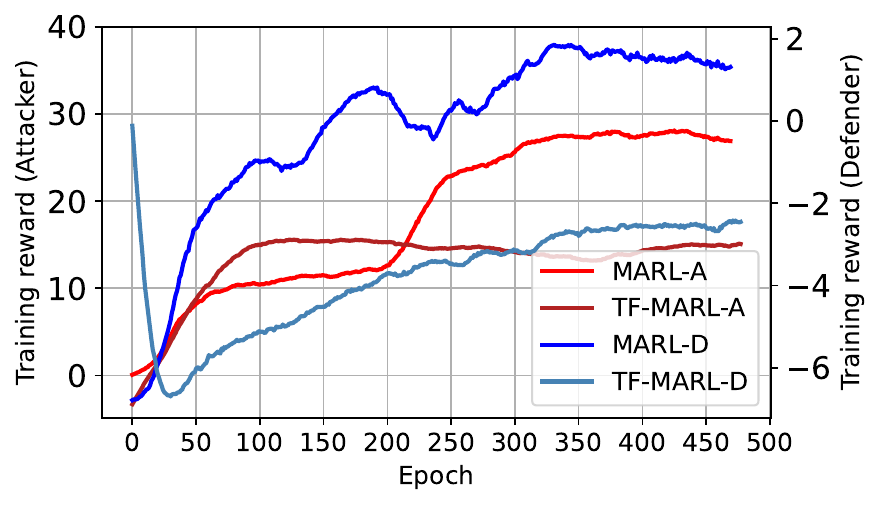}
    \caption{History of the agent's rewards as a function of epoch in the frequency regulation case study. The solid line is the mean value of the rewards obtained with five independent training runs.}
    \label{fig:train_fre}
\end{figure}

Compared to voltages, frequency dynamics exhibit greater variability, thereby posing challenges for the learning process and resulting in lower detection accuracy compared to Case Study I. As shown in Fig.~\ref{fig:train_fre}, the TF-MARL-D retains expert knowledge and achieves promising performance in the initial training stage. However, the reward first decreases when TF-MARL-A launches attacks that evade detection, and subsequently increases as TF-MARL-D adjusts its control policy to defend against the attacker. The final reward of TF-MARL-D is smaller than that of MARL-D, possibly because the warm-up strategy encouraged the adversary to explore less impactful and more stealthy FDIA. Table \ref{tab:p_com_wup_fre} shows that TF-MARL-D consistently outperforms the offline defender when exposed to the TF-MARL-A adversary (performance increases between 40\% and 225\% across $5$ independent runs). Compared to MARL-A (Table \ref{tab:p_com_nowup_fre} first row), the frequency deviation induced by TF-MARL-A is reduced by MARL-A. 

\begin{table}[ht]
\begin{center}
\renewcommand{\arraystretch}{1.1}
\caption{Detection accuracy (\%) under TF-MARL-A's generated attacks under different $c_\text{wup}$ setup values in the frequency regulation case study}
\label{tab:p_com_wup_fre}
\begin{tabular}{ |c|c|c|c|c|c|c||c| } 
\hline
\multirow{3}{*}{ $100$} & $\sum_{t \in \mathcal{T}}r_t^w$ & 20  & 19 & 21 & 22 & 18 & Mean
\\
\cline{2-8}
& TF-MARL-D & 33\% & 39\% & 39\% & 42\% & 46\% & \textbf{40\%}
\\
\cline{2-8}
& Offline-D & 22\% & 20\% & 12\% & 19\% & 33\% & 21\%
\\
\hline
\hline
\multirow{3}{*}{$1$} & $\sum_{t \in \mathcal{T}}r_t^w$ & 16 & 16 & 17 & 19 & 17 & Mean
\\
\cline{2-8}
& TF-MARL-D & 28\%  & 29\% & 29\% & 31\% & 33\% & \textbf{30\%}
\\
\cline{2-8}
& Offline-D & 19\%  & 25\% & 20\% & 18\% & 20\% & 20\%
\\
\hline
\end{tabular}
\end{center}
\end{table}

\begin{figure}[ht!]
    \centering
    \includegraphics[width=1\linewidth]{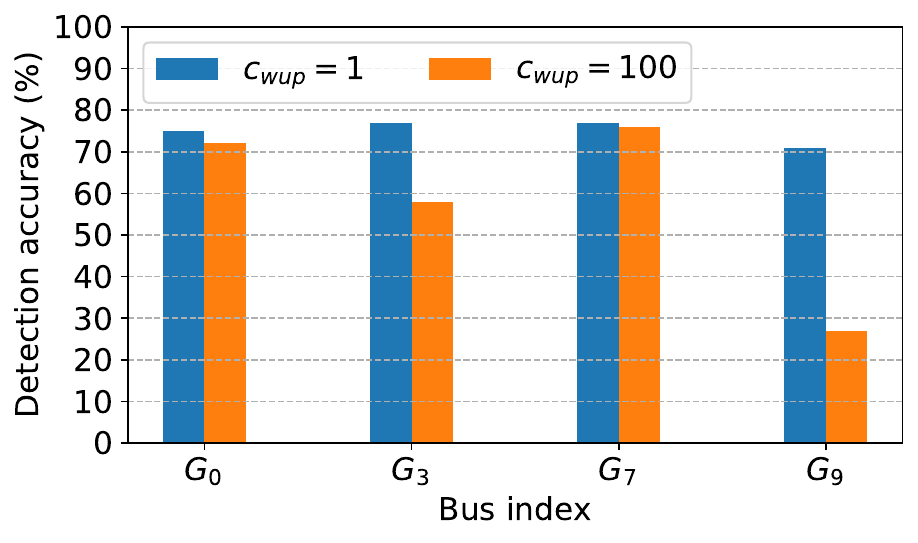}
    \caption{The detection accuracy of TF-MARL-D under time-invariant attacks in the frequency regulation case study.}
    \label{fig:p_com_invi_coe}
\end{figure}

Given the importance of transfer learning to retain prior knowledge, an additional investigation was conducted to determine how to best conduct transfer learning. Here, the importance of the prior data is constrained with $c_{\text{wup}}$. Smaller $c_{\text{wup}}$ leads to larger model weight values, which leads to performance degradation in defending newly generated FDIAs during online training. As shown in Fig. \ref{fig:p_com_invi}, when $c_\text{wup} = 100$, the detection accuracy of TF-MARL-D at $G_9$ degrades significantly. This could be because MARL-A is unlikely to attack $G_9$ during training, which leads to minimum frequency deviations compared to other inverters \cite{Romesh}. As shown in Fig. \ref{fig:p_com_invi_coe}, a smaller $c_\text{wup}$ can help retain more knowledge and achieve higher accuracy attacks occurring at $G_9$. As shown in Table \ref{tab:p_com_wup_fre}, when $c_{\text{wup}}$ decreases from 100 to 1, the detection accuracy drops from $40\%$ to $30\%$. Thus, we properly tune $c_{\text{wup}}$ and adopt 100 to avoid the weight domination but still enable the TF-MARL-D to remember the Offline-D's weights to defend against time-invariant attacks. 



\section{Conclusions}\label{sec:conclusion}
In this work, an MARL framework is proposed to craft a data-based defense capable against impactful FDIA that were not in the original dataset. Compared to an offline defender, the MARL defender can continuously learn to defend against newly generated adversarial FDIAs and adjust its defense strategy promptly. Thanks to concurrent training, the defender and the adversary are exposed to a diverse set of environments but that is not sufficient to ensure that a defender can catch all possible FDIAs. In particular, synthetically generated FDIA was shown to escape the MARL defender. Instead, it is proposed to make sure that the defender is accurate against synthetic and adversarially generated FDIA through a transfer learning approach. In that way, it is shown that a data-based defender can still be accurate against a wide range of FDIAs in both distribution and transmission systems. 

In this work, we retain prior knowledge by warm-starting the weights, which may sacrifice the generalization ability of the defender if weight domination occurs. This means that the MARL defender cannot change too much to avoid forgetting. A cure would be show both synthetic and adversarial data to the defender at training time which is left for future work.

\bibliographystyle{IEEEtran}
\bibliography{refs}

\end{document}